\documentstyle[aps,multicol,epsfig]{revtex}
\newcommand{\nc}{\newcommand}
\nc{\be}{\begin{equation}}
\nc{\ee}{\end{equation}}
\nc{\bea}{\begin{eqnarray}}
\nc{\eea}{\end{eqnarray}}
\nc{\bean}{\begin{eqnarray*}}
\nc{\eean}{\end{eqnarray*}}
\nc{\mb}{\mbox}
\nc{\rnc}{\renewcommand}
\nc{\r}{\mb{\boldmath$r$}}
\nc{\x}{\mb{\boldmath$x$}}
\nc{\A}{\mb{\boldmath$A$}}
\nc{\sa}{\mb{\boldmath$a$}}
\nc{\nab}{\nabla}
\nc{\X}{\sf x}

\begin{document}
\draft

\def\del{\partial}

\title{ Strong quasi-particle tunneling study in the paired quantum Hall states }
\author
{Kentaro Nomura and Daijiro Yoshioka
 }

\vspace{5mm}

\address{
Department of Basic Science, University of Tokyo, 3-8-1 Komaba, Tokyo 153-8902\\
}

\vspace{4mm}

\date{\today}
\maketitle

\vspace{0.5cm}

\begin{abstract} 
  The quasi-particle tunneling phenomena in the paired 
fractional quantum Hall states
are studied. A single point-contact 
system is first considered. Because of relevancy
of the  quasi-particle tunneling term, the
strong tunneling regime should be investigated.
 Using the instanton method 
it is shown that 
 the strong quasi-particle tunneling 
regime is described
as the weak electron tunneling regime effectively. 
  Expanding to the network model the paired quantum Hall liquid to insulator 
transition is discussed.
\end{abstract}

\vspace{0.5cm}



\begin{multicols}{2}
\narrowtext

{\bf {1. Introduction}}

\vspace{3mm}

The integer and fractional quantum Hall effect was observed 
in a two-dimensional electron
systems subjected to a strong perpendiclar magnetic field.
\cite{iqhe,fqhe}
In these experiments, the Hall resitivity showed plateau behavior
and the longitudinal resitivity vanished in odd denominator filling
of the Landau level. \cite{smg}
 Contrary to the odd denominator filling fractions,
 the quantum Hall effect doesn't appear in $\nu =1/2$.
\cite{jiang}
This state is regarded as Fermi liquid state of the composite
fermions which have charge $e$ and the flux $2\phi_0\ (\phi_0=
hc/e)$.\cite{jain}
At mean field level, the fictitious fluxes attached to composite 
fermions  cancel the external magnetic field and the fermions
system is described as that in the absence of the 
magnetic field. \cite{hlr}
On the other hand, the even denominator 
quantum Hall effect was observed in single layer
 systems at $\nu=5/2$, \cite{will} and
  in the double layer systems at 
 $\nu=1/2$.\cite{eis}
A large number of theoretical studies have been made on 
the grand state of the even denominator quantum Hall systems. 
To date three possible states have been proposed
 to explain such  incompressible 
regimes at even denominator filling factor. 
They are Halperin(331),\cite{331} Moore-Read(Pfaffian),\cite{mr}
 Haldane-Rezayi states. \cite{hr}
 These states are regarded as the BCS states 
in the composite fermion theory.
The order parameter symmetry is triplet p-wave 
with ABM and A$_1$ type in
 $^3$He for 331 and Pfaffian state respectively, 
and d-wave for Haldane-Rezayi state.
\cite{gww,ho,morinari,rg}
  The wave functions of the grand state and the
 excited states for
 these paired states are represented as the conformal 
block of proper conformal field theories. 
In the case of the Pfaffian, 331, and 
Haldane-Rezayi states, corresponding theories 
have a central charge
$c= 1+1/2, 1+1, 1-2$, respectively. \cite{mr,mir,gfn,lw}
These conformal theories also
 describe the edge excitations on 
the boundaries of the sample. 

  Edge quasi-particle tunneling is vital
 to study the quantum transport
phenomena in the quantum Hall systems.
   Tunneling experiments in the point-contact systems 
have played a central role for detecting the non-Fermi
liquid properties of fractional quantum Hall liquids.
   Milliken et al. studied the transmission through 
a point contact in a gated structure. \cite{mil}
At $\nu=1/3$ fractional quantum Hall plateau regime,
 the temperature dependence of the two-terminal 
conductance is non-linear
$G \propto T^{\alpha}$ where $\alpha =4$ for  $\nu=1/3$. 
Another important  experiment is the measurement
 of the shot noise which can measure the fractional charge. 
       In such  experiments, 
the quasi-particle tunneling between the edges is substantial.
This can be described by a Tomonaga-Luttinger liquid
 model with a scattering
potential at $x=0$. \cite{kf,fn}
Many theorists have studied these 
effects and expanded into more
complex situations. \cite{imuranag,imuraino}     
  In recent work, we also have carried this problems to the
hierarchical states and explained Reznikov et al's experiment.
 \cite{reznikov,imuranomura}
 In addition, we have constructed an edge state network model for
the hierarchical state and discussed the transition 
from a quantum Hall
liquid to another quantum Hall liquid or an insulator. 
\cite{nomurayoshioka}  
   In this paper we study quasi-particle tunneling phenomena in the paired fractional
quantum Hall states. 
      Because of relevancy in the renormalization group analysis, 
we have to study quasi-particle tunneling 
 non-perturbatively as well as the Laughlin state and the 
hierarchical state. 
Using the instanton method we show that in the strong quasi-particle tunneling regime 
quasi-particle tunneling is described as weak electron tunneling
 effectively in the Pfaffian, and
331 states.  These results are very similar to the Laughlin and hierarchical
states. We introduce the edge state network model and discuss 
the paired quantum Hall liquid to insulator transition.
      
 
{\bf {2. Quasi-hole tunneling model}}

     
  We consider the model which describes a two-terminal
 Hall Bar geometry
where a two-dimensional electron system between the left and 
right terminals has upper
and lower edges with a scattering at $x=0$. 
 We start to analyze the $\nu=1/q$ generalized Pfaffian state.
 The $c=1/2$ part in the edge modes belong to the
 same universality
class as the critical point of the two-dimensional Ising model
 which is
described by the real fermion field $\psi$ and the 
spin operator $\sigma$.\cite{mr}
These modes are interpreted as the pair breaking 
and the half quantum vortex excitation.
 On the other hand $c=1$ part describes the Laughlin type
excitations which corresponds to the free boson theory.\cite{wen1,wen2}
So, the edge state theory is written as

\bea
  S_0 &=& 
\int {\rm d}\tau {\rm d}x 
\frac{1}{8\pi \nu}
\left[
\frac{1}{v} \left( \frac{\partial \phi}{\partial \tau} \right)^2 
+v\left( \frac{\partial \phi}{\partial x} \right)^2 
\right] \nonumber \\
&&\ \ \ \ \ +
{\overline {\psi}} 
\left( \frac{\partial }{\partial \tau} -{\rm i}v
\frac{\partial }{\partial x}
 \right) 
{\overline {\psi}} +
\psi 
\left( \frac{\partial }{\partial \tau} +{\rm i}v
\frac{\partial }{\partial x}
 \right) 
\psi. 
\eea
The electron, and the quasi-hole operators are given by
$\psi_e (z) = \psi \ e^{-{\rm i}\phi / \nu} \  \
 \psi_{qp}(z)= \sigma \ e^{-{\rm i}\phi/2},$ 
then the quasi-particle, and electron tunneling
 Hammiltonian
is written as
\bea
 H_{\sf QPT}&=& t_{\sf QPT}
 (\psi_{\rm qpR}^{\dag} \psi_{\rm qpL}
  + {\sf h.c.})\nonumber     \\
&=& \Gamma_{\sf QPT} \sigma(x=0) \cos(\phi(x=0)/2), \\
 H_{\sf ET}&=& t_{\sf ET}
 (\psi_{\rm eR}^{\dag} \psi_{\rm eL}
  + {\sf h.c.})\nonumber \\
&=& \Gamma_{\sf ET} {\rm i}\psi {\overline {\psi}}
\cos(\phi(x=0)/\nu)
\eea
respectively, where $t_{\sf QPT}(t_{\sf ET})$ is the tunneling 
amplitude of the quasi-particle (electron). 
We set the value of
 $t_{\sf QPT}$ positive.

Counting the conformal dimension
we can derive the renormalization group
 equation for quasi-particle tunneling as
\bea
 \frac{{\rm d} {\Gamma}_{\sf QPT}}{{\rm d} l} = \left[   1-\left( 
\frac{\nu}{4}+ \frac{1}{8}
\right)  \right]  {\Gamma}_{\sf QPT},
\label{RGeq1}
\eea
where ${\rm e}^{l}=\frac{\Lambda_0}{\Lambda'}$, $\Lambda_0$ and
$\Lambda'$ are bare, renormalized cut-off, respectively.
The $\beta$-function (right hand side of (\ref{RGeq1}))
 is positive for any $\nu \leq 1$,
 so we have to study the strong coupling regime non-perturbatively.

\vspace{3.0mm}

{\bf {3. Instanton method}}

\vspace{3mm}

To study this problem, we introduce an effective theory for the 
non-linear degree of freedom $\phi(x=0)\equiv \theta$ in (2).
as following;


\bea
Z&=& \int {\cal D}\psi {\cal D}{\overline {\psi}} {\cal D}\phi 
\nonumber \\
&& \ \ \ \ \ \ \ \ \ 
\ \exp \left( -S_{\rm B}-S_{\rm F}-
\Gamma_{\sf QPT} \int {\sf d}\tau \ \sigma \cos \phi/2 
 \right)\nonumber \\
&=& \int {\cal D}\psi {\cal D}{\overline {\psi}} 
e^{-S_{\rm@F}}\int{\cal D}\phi
{\cal D}\theta\
 \delta\left(  \theta-\phi(x=0)   \right) 
\nonumber  
\\
&&\ \ \ \ \ \ \ \ \ 
\exp{\left(-S_{\rm B}-
\Gamma_{\sf QPT} \int {\sf d}\tau \ \sigma \cos \theta/2  \right)}
  \nonumber \\
&=& \int {\cal D}\psi {\cal D}{\overline {\psi}} {\rm e}^{-S_{\rm F}}\int
 {\cal D}\phi
{\cal D}\theta    
\int   {\cal D}\lambda\   
{\rm e}^{-\int {\rm d} \tau \ \lambda (\theta- \phi(x=0) ) }
 \nonumber 
\\
&&\ \ \ \ \ \ \ \ \ 
\exp{\left(-S_{\rm B}- \Gamma_{\sf QPT} 
\int {\sf d}\tau \ \sigma \cos \theta/2  
  \right)}
\eea
where $S_{\rm F}$ and $S_{\rm B}$ are the action for the real fermion and
boson respectively. Integrating out $\phi$ and $\lambda$, we have;
\bea 
Z&=& \int {\cal D}\psi {\cal D}{\overline {\psi}}{\rm e}^{-S_{\rm F}}\int
 {\cal D}\theta  \nonumber \\ && \exp     
{\left(-\sum_{\omega} \frac{|\omega |}{4\pi \nu}
\theta(-\omega) \theta(\omega)
 -\Gamma_{\sf QPT} 
\int {\sf d}\tau \ \sigma \cos \theta/2  
 \right)}. \nonumber \\
\label{z}
\eea
In the view point of the dynamics for $\theta$, 
the first term in (\ref{z}) is
the friction term in the Caldeira-Legget theory, \cite{cl}
and the second term is regarded as the periodic potential.
 On the other hand, in terms of the Ising model, 
the second term in (\ref{z}) is 
interpreted as the "magnetic field" which
is proportioned to $\cos \theta/2$.
 In the strong coupling limit, due to the "magnetic field" 
the saddle point solution of the spin field
$\sigma(x=0)$ is no longer zero. If $\sigma <0$ the $\theta$
has a value $4\pi n$, If $\sigma >0$ $\theta= 2\pi(2n+1)$, 
where $n$ is an integer.
 The excitations for the strong coupling limit  are described
by the instanton solutions  same as the Laughlin state or the 
problems of  single impurity in non-chiral Luttinger liquid.
 In this situation, however, there is an {\it {intrinsic difference 
}} with them.
It is the degree of freedom for the spin operator $\sigma$.
 Specifically usual instanton solutions (we say it 
"standard instanton") are defined as a step from a minimum to another
 minimum of the potential.  
 The other process carry {\it {spin flip}} 
which causes the reversal of 
the sign  of the potential, 
so the minima and maxima are interchanged.
We say the latter solution "half instanton".
In the strong coupling regime, the latter
 contributions are dominant.
  We have to note that 
in the half instanton process the dynamics 
of Ising systems should be considered.

Using the instanton method
we construct an effective action for the strong coupling regime.
We introduce a function defined as 
\bea
 \frac{{\sf d}\theta_{\sf ins}}{{\sf d}\tau} \equiv h(\tau),
\eea
where $\theta_{\sf ins}$ is the solution for  single half instanton.
So the differential of the  $n$ instanton solution and its Fourier
coefficient are  written as 
\bea
 \frac{{\sf d}\theta_n}{{\sf d}\tau} &=&\sum_{i=1}^n e_i h(\tau-\tau_i) \\
 -{\rm i}\omega \theta_n(\omega)&=&
 \sum_{i=1}^n e_i {\tilde h}(\omega)
 e^{{\rm i} \omega \tau_i},
\eea   
where $e_i$ is the charge of the instantons,
and ${\tilde h}(\omega)$ is the Fourier coefficient of $h(\tau)$.
Especially the value;
\bea
 {\tilde h}(\omega =0) &=&\frac{1}{\sqrt {\beta}}
 \int d\tau h(\tau) 
  = \frac{2\pi}{\sqrt {\beta}} 
\eea
will be important.
 To derive the effective action, we consider a grand canonical
ensemble of the half instantons.
The  contributions of the standard instantons are not important in
the strong coupling regime. We neglect the $\omega$-dependence
of $h(\omega)$ in the first term in (\ref{z}) as

\bea
 S_D &\equiv& \sum_{\omega} \frac{|\omega|}{4\pi \nu}
\theta(-\omega)\theta(\omega)  \nonumber \\
&\cong& \sum_{ij}^n 
\left(    
\frac{\pi  }{\nu}\frac{1}{\beta} \sum_{\omega}
\frac{1}{|\omega|} e^{-{\rm i}\omega(\tau_i-\tau_j)}
 \right) e_i e_j.
\eea
The grand canonical partition function is given as 
\bea
  Z &=& \int {\cal D}\psi {\cal D}{\overline {\psi}} e^{-S_F}
\sum_{n=0}^{\infty} \sum_{\{e_i\}}
\nonumber \\
&&
\int_0^{\beta} d\tau_n \int_0^{\tau_n} d\tau_{n-1}
\cdots \int_0^{\tau_2} d\tau_1 \prod_{i=0}^{n} z(\tau_i)
\nonumber \\  &&
 \exp{\left[ 
-\sum_{ij}^n \left(    
\frac{\pi  }{\nu}\frac{1}{\beta} \sum_{\omega}
\frac{e^{-{\rm i}\omega(\tau_i-\tau_j)}}{|\omega|} 
 \right) e_i e_j
  \right]  }
\eea
where 
$z(\tau) = \omega_0 e^{-S_{\sf ins}}$
is  the fugacity of the instanton, $\omega_0$ is a characteristic
energy of the tunneling process. The value must be directly 
proportional to the energy operator $\varepsilon$,
$i.e.\ \omega_0 =z_0 \varepsilon(\tau)$. 
Where $z_0$ is a constant.
Introducing 
Stratonovich-Hubberd field $\theta$ we get
\bea
Z
&=& \int {\cal D}\psi {\cal D}{\overline {\psi}} e^{-S_F}
\sum_{n=0}^{\infty} \sum_{\{e_i\}} \frac{1}{n!}
\nonumber \\ &&
\int_0^{\beta} d\tau_n \int_0^{\beta} d\tau_{n-1}
\cdots \int_0^{\beta} d\tau_1 
\ z_0^n \prod \varepsilon(\tau_i)
\nonumber \\ &&\ \ 
 \int {\cal D}\theta  \nonumber 
\exp{\left[ -  \sum_{\omega} \frac{|\omega|}{4\pi \nu}
\theta(-\omega)\theta(\omega)
+ \frac{\rm i}{\nu} \sum_i e_i \theta(\tau_i)  
 \right]  }   \nonumber \\
&=&  \int {\cal D}\psi {\cal D}{\overline {\psi}} e^{-S_F}
\int {\cal D}\theta \exp \nonumber   
[
   -\sum_{\omega} \frac{|\omega|}{4\pi \nu}
\theta(-\omega)\theta(\omega) 
\nonumber \\
&&\ \ \ \ \ \ \ \ \
\ \ \ \ \ \ \ \ \ \
 - 2 z_0\int d\tau \ {\rm i}\psi{\overline {\psi}}
 \cos{( \theta(\tau)
  /\nu  )} 
 ]. 
\eea
Note the second term is the electron tunneling term (see eq.(3)),
and $z_0$ is proportional to the tunneling amplitude of the electrons.
As the case of the Laughlin state, we can treat the quasi-hole
tunneling as the weak electron tunneling effectively
 in the strong coupling regime.
The RG-equation for weak electron tunneling is given as
\bea
 \frac{{\sf d} z_0}{{\sf d} l} = \left[   1-\left( 
\frac{1}{\nu} + 1
\right)  \right] z_0,
\eea
and the dependence of the two-terminal conductance is
\bea
  G\ \propto \ z_0^2
(\Lambda_0)\ T^{
2/ \nu },
\eea
where $\Lambda_0$ is a bare cut-off.

 Next we consider the case of the 331 state. 
The edge modes theory is written as;
\begin{eqnarray}
S_0 &=& \int {\rm d}\tau {\rm d}x \  \sum_{IJ} \frac{{\rm i}}{4\pi} K_{IJ}
  \frac{\partial \phi_I^+}{\partial \tau}
  \frac{\partial \phi_J^-}{\partial x}       
   \nonumber     \\
& &\ \ \ \ \ \ \ 
   + \frac{1}{8\pi}v_{IJ}( \frac{\partial \phi_I^+}{\partial x}
\frac{\partial \phi_J^+}{\partial x} 
+ \frac{\partial \phi_I^-}{\partial x}
\frac{\partial \phi_J^-}{\partial x} ), 
\end{eqnarray}
\bea
 K = \left(
  \begin{array}{@{\,}cc@{\,}}
   3 & 1 \\
   1 & 3
 \end{array}
 \right) ,
\ \ \ \  t = \left(
    \begin{array}{@{\,}cc@{\,}}
         1 \\
         1
     \end{array}
     \right).
\eea
In the charge-spin basis, $\phi_{c,s}=\phi_{\uparrow}\pm \phi_{\downarrow}$,
the two component can be separated as
\begin{eqnarray}
{\cal L}_0   &=& (\  
\frac{{\rm i}}{4{\pi} K_{c}}\frac{\partial \phi_c^+}{\partial \tau}
\frac{\partial \phi_c^-}{\partial x} + 
\frac{v_c}{8\pi K_{c}}[(\frac{\partial \phi_c^+}{\partial x})^2
 + (\frac{\partial \phi_c^-}{\partial x})^2]   \nonumber \\
&&\ \ \ \   + \frac{{\rm i}}{4{\pi} K_{s}}
\frac{\partial \phi_s^+}{\partial \tau}
\frac{\partial \phi_s^-}{\partial x} + \frac{v_s}{8\pi
 K_{s}}
[(\frac{\partial \phi_s^+}{\partial x})^2 + 
(\frac{\partial \phi_s^-}{\partial x})^2]\  ),  
  \nonumber \\
\end{eqnarray}
where $K_c=1/2=\nu, \ K_s=1$.

The electron, quasi-hole operators are given by
\bea
 \psi_{e\uparrow}= e^{-3i\phi_{\uparrow}-i\phi_{\downarrow}}, 
 \psi_{e\downarrow}= e^{-i\phi_{\uparrow}-3i\phi_{\downarrow}} \\
  \psi_{qp\uparrow}= e^{-i\phi_{\uparrow}}, 
 \psi_{qp\downarrow}= e^{-i\phi_{\downarrow}} 
\eea
Then the quasi-hole and the electron tunneling Hamiltonian is written as
\bea
 H_{\rm QPT} &=& \Gamma_{\rm QPT}\ \sum_{\sigma=\uparrow, 
\downarrow} 
\cos\phi_{\sigma} \nonumber \\
  &=& 2 \Gamma_{\rm QPT}\ \cos\frac{\phi_c}{2} 
\cos\frac{\phi_s}{2}\\
 H_{\rm ET} &=& 2\Gamma_{\rm ET}\  \cos 2\phi_{c} \cos\phi_s
\eea
We derive
the effective action for $\theta_c=\phi_c(x=0)$ 
and $\theta_s=\phi_s(x=0)$ like eq.(6) 
\bea
 S &=& \sum_{\omega} \frac{|\omega|}{4\pi K_c} \theta_c(-\omega)
\theta_c(\omega)
+ \sum_{\omega} \frac{|\omega|}{4\pi K_s} 
\theta_s(-\omega)
\theta_s(\omega) \nonumber \\
&&\ \ \ \ \ \ \ \ \ \ + \int d\tau \  2\Gamma_{\rm QPT}\
 \cos\frac{\theta_c}{2} \cos\frac{\theta_s}{2}
\eea
In the present case, the  instantons are specified by 
the change of the values of the phase 
$\phi_c, \phi_s$. 
 Most important type in the strong coupling regime is the $(\pm \pi,\pm \pi)$.
Considering the grand-canonical ensemble and introducing the Stratonovich-Hubbard field
we can construct the effective action;
\bea
 S' &=& \sum_{\omega} \frac{|\omega|}{4\pi K_c} \theta_c(-\omega)
\theta_c(\omega)
+ \sum_{\omega} \frac{|\omega|}{4\pi K_s} \theta_s(-\omega)
\theta_s(\omega) \nonumber \\
&&\ \ \ \ \ \ \ \ \ \ + \int d\tau \ 2\Gamma_{\rm ET}\
 \cos2\theta_c \cos\theta_s.
\eea
Note the last term of (24) is the electron tunneling term again.
Therefore in the case of the 331 the quasi-hole tunneling also can be regarded
as weak electron tunneling.
Temprature dependence of the two-terminal conductance can be 
calculated as (15), and we have  $G \propto T^4$.

\vspace{3mm}
{\bf {4. Paired quantum Hall liquid to Hall insulator transition}}
\vspace{3mm}

We consider the situation to increase the magnetic field. 
Increasing the strength
the longitudinal resistance arises and the system  becomes 
the Hall insulator.
Near the critical point, the phase separation should occur.
In  such a condition, the edge state network model
 is applicable.
\cite{simshoni,pryadko,nomurayoshioka}
In previous work, to discuss the quantum Hall plateau-plateau 
transition and 
quantum Hall liquid to insulator transition,
 we introduced a hierarchical filling version of 
the  edge state network model.
The model has the 
dual description: weak quasi-particle tunneling regime can be regarded as
the (hierarchical) quantum Hall liquid phase, 
while the strong quasi-hole tunneling 
regime (weak electron tunneling regime) corresponds
 to the insulating phase.
 Because the former case corresponds to the situation
 that there are
 some small vortices condensed regions in the sea of filled
 quantum Hall liquid, quasi-particle tunneling occurs
between edges which enclose the vortices condensed regions.
On the other hand the latter case corresponds to the opposite 
situation. Namely there are some small quantum Hall droplets
connected by  electron tunneling where the vortices condensate 
regions are dominant.
If one increases the strength of randomness,
 the liquid-insulator transition occurs also.
\cite{nomurayoshioka}
 We can apply the model to the pairing state. 
Reconstructing the instanton method the dual symmetry has been shown.
As well as the Laughlin and hierarchical version, the strong quasi-hole tunneling
regime should be regarded as the Hall insulator.
The dual symmetry found by us in present paper suggests $I-V$ 
reflection symmetry which is observed in 
Shahar's experiment near the critical point of
  the $\nu=1/3$ fractional quantum Hall liquid-insulator 
transition.
However, even though in the case of $\nu=1/3$,
 no one could show rigidly the fact that
if at $B>B_c$ $I-V$ character is given by $I= F(V)$, 
at $B<B_c$ $I=F^{-1}(V)$.
This point is left to a future problem. 
 
 Finally we note the transition  from a 331 state to 
another state in the $\nu=1/2$
double layer quantum Hall systems.
In eq.(21) we treated the tunneling amplitude symmetrically for the spin index
as $\Gamma_{{\rm QPT} \uparrow}=\Gamma_{{\rm QPT} \downarrow}$.
However, in the $\nu=1/2$ double layer quantum Hall systems,
 the values of them 
may be different.  
If only one of the coupling constant
(say it $\Gamma_{{\rm QPT} \downarrow}$)
  become large and the other $\Gamma_{{\rm QPT} \uparrow}$ 
left small, 
the $\nu=1/2$ plateau to $\nu=1/3$
plateau transition can  occur  
for the systems without inter-layer tunneling.
Because
the field $\phi_{\downarrow}$ is pinned at the tunneling 
points, and it cannot contribute to the transport,
 the effective action for $\phi_{\uparrow}$ which should respond to the
 electric perturbation is written as
that of at $\nu_{\sf eff}=1/3$. \cite{imuranomura,nomurayoshioka}

In this paper we have investigated quasi-particle 
tunneling. We showed that  
we can treat the strong
quasi-particle tunneling regime as weak electron tunneling.
This dual symmetry  hold in the edge state network model.
We insist the possibility of observation of 
the $I-V$ reflection symmetry
near the liquid-insulator transition.

\vspace{2mm}
{\bf Acknowledgements}
\vspace{2mm}

  We are grateful to K. Imura
 for the useful discussion.
This work is supported by a Grant-in-Aid for Scientific Research (C) 10640301
from the Ministry of Education, Science, Sports and Culture.

\end{multicols}
\end{document}